\documentclass[12pt]{iopart}

\usepackage{iopams}  
\usepackage{graphicx}
\usepackage{dcolumn}
\usepackage{bm}
\usepackage{subfigure}
\usepackage{psfrag}
\usepackage{color}
\newcommand{\text}[1]{\textrm{#1}}

\begin{document}

\title[3D Full dynamic of finite size particles together with the carrier turbulent flow field.]{Simultaneous 3D measurement of the translation and rotation of finite size particles and the flow field in a fully developed turbulent water flow.}

\author{Simon Klein$^{1}$}%
\author{Mathieu Gibert$^{1,4,5}$}
\author{Antoine B\'erut$^{1}$}%
\author{Eberhard Bodenschatz$^{1,2,3,4}$}%
\address{ 
$^{1}$Max Planck Institute for Dynamics and Self Organization, D-37073 G\"ottingen, Germany
}%
\address{
$^{2}$Institute for Nonlinear Dynamics, University of G\"ottingen, D-37073 G\"ottingen, Germany
}%
\address{
$^{3}$Laboratory of Atomic and Solid-State Physics and Sibley School of Mechanical and Aerospace Engineering, Cornell University, Ithaca, New York 14853
}%
\address{
$^{4}$International Collaboration for Turbulence Research
}%
\address{
$^{5}$Now at : Institut N\'eel, CNRS/UJF, F-38042 Grenoble Cedex 9, France
}%

\ead{mathieu.gibert@grenoble.cnrs.fr}
\begin{abstract}
We report a novel experimental technique that measures simultaneously in three dimensions the trajectories, the  translation, and the rotation of finite size inertial particles together with the turbulent flow. The flow field is analyzed by tracking the temporal evolution of small fluorescent tracer particles.  The inertial particles consist of a super-absorbent polymer that renders them index  and density matched with water and thus invisible.  The particles are marked by inserting at various locations tracer particles into the polymer. Translation and rotation, as well as the flow field around the particle  are recovered dynamically from the analysis of the marker and tracer particle trajectories. We apply this technique to study the dynamics of inertial particles much larger in size ($R_{p}/\eta \approx 100$) than the Kolmogorov length scale $\eta$  in a  von K\'arm\'an swirling water flow ($R_{\lambda}\approx 400$). We show, using the mixed (particle/fluid) Eulerian second order velocity structure function, that the interaction zone between the particle and the flow develops in a spherical shell of width $2R_{p}$ around the particle of radius $R_{p}$.  This  we interpret as an  indication of a wake induced by the particle. This measurement technique has many additional advantages that will  make it useful to address other problems such as particle collisions, dynamics of non-spherical solid objects,  or even of wet granular matter.

\end{abstract}

\vspace{2pc}
\submitto{\MST}
\maketitle

\section{Introduction}
Particle-laden flows  are prevalent in natural and technological flows.  For example they are relevant to warm cloud dynamics~\cite{falkovich:2002,Shaw:2003p769} with impact on the  climate,  or to dust and pollution transport in atmosphere and oceans,  or can be found in most  technological  processes where matter is mixed in fluids.  In the past few years, significant advances have been made in experimental approaches to particle dynamics in turbulent flows thanks to the rapid development of visualization based three dimensional measurement techniques such as Lagrangian Particle Tracking (LPT)~\cite{Ouellette:2006p384,Xu:2008p1921}.  When studying the fluid's dynamics with LPT, one has to assure that the particles behave passively and do not perturb the flow. This goal can be considered reached when the particles are density matched to the fluid, which avoids buoyant and inertial forces, and when their sizes are much smaller than the smallest length scale of the velocity gradients in the flow. Such particles are then called \textit{tracer particles}. These requirements are very stringent and can be hard to achieve in some flows, {\it{e.g.}} thermal turbulent convection, where fluid density changes strongly in the thermal boundary layer~\cite{2008JFM...605...79S,Gibert:2009jv,Gasteuil:2007de}; highly turbulent flows, which exhibit very fast dynamics (fractions of milliseconds) on length scales that reach micrometers~\cite{Voth:2002p432} (indeed $L/\eta\propto Re^{3/4}$, where $L$ is the large scale of the flow, $\eta$ is the characteristic length-scale of the smallest eddy in the flow, also called the Kolmogorov length scale, and $Re$ is the Reynolds number of the flow);  or in the study of superfluid turbulence where the normal and superfluid coexist~\cite{Bewley:2008hn}.  When the  density of the particles is different from that of the fluid, or the particle size is large, the particles  cannot  be regarded as tracers of the fluid and are called \textit{inertial particles}.\\

In this paper, we focus on the study of large particles in turbulent flows, \textit{i.e.} particles much larger in size than the Kolmogorov length scale $\eta$ of the flow. The particles that we consider here have a radius $R_{p}\approx 100\eta\approx 0.1L$.
 Experiments on two-particle statistics of  small and heavy particles can be found in \cite{EPL_particle_2010,JFM_particle_2012}. Most theoretical and numerical simulations of the complex coupling between the motion of \textit{inertial particles}  and their carrier fluid rely on simplified equations such as the Maxey-Riley-Gatignol equation~\cite{MaxeyRiley,GATIGNOL:1983p3869}. This equation is only valid for point-like particles and couples the motion of the particle to the fluid, without any back-reaction of the particle motion on the flow (one-way coupling). This approach has been proven to be sufficient to describe the rich dynamics of small and heavy particles~\cite{PartRev09}, but has also found its limit~\cite{EPL_particle_2010,Qureshi:2007p1316,Qureshi:2008p2080,Volk:2008p271,Calzavarini:2009de,bubble_Lohse_2012} (even when corrected by the so called Fax\'en corrections) when considering particles larger than $4\eta$. To go beyond this simplified model, Hohmann and Bec ~\cite{Homann:submitted} developed a direct numerical simulation using a dynamical pseudo-penalisation technique in order to satisfy the non-slip boundary conditions at the surface of a unique particle evolving in a turbulent flow. This strategy is very promising (as the one developed in \cite{Naso:2010p4267}, in which the particle is static) but also comes with high computational cost that currently restricts it to very low Reynolds numbers ($R_{\lambda}^{max}=72$ for~\cite{Homann:submitted} and $R_{\lambda}^{max}=20$ for~\cite{Naso:2010p4267}, where $R_{\lambda}$ is the Reynolds number based on the Taylor micro-scale of the flow). Therefore, for the foreseeable future, the investigations on large particle dynamics in highly turbulent flows  must rely heavily on experimental work. 
Few experiments such as \cite{Ouellette:2008ds} have been conducted in 2D flows, also 2D studies of 3D flows based on Particle Imaging Velocimetry can be found in~\cite{POELMA_2007}, but a complete experimental 3D approach resolving the particles dynamics and the flow has to our knowledge not been reported.
The recent work of Zimmermann~\textit{et al.}~\cite{2011PhRvL.106o4501Z,2011RScI...82c3906Z} has been able to resolve the full motion (translation and rotation) of single  finite-sized particles in a highly turbulent flow. Here, we present a technique that  goes beyond and  resolves simultaneously the full three dimensional motion of the particles together with the turbulent flow field around it.

\section{Experimental setup}
\begin{figure}
\begin{center}
  \subfigure[Side view]{\includegraphics[width=0.57\textwidth]{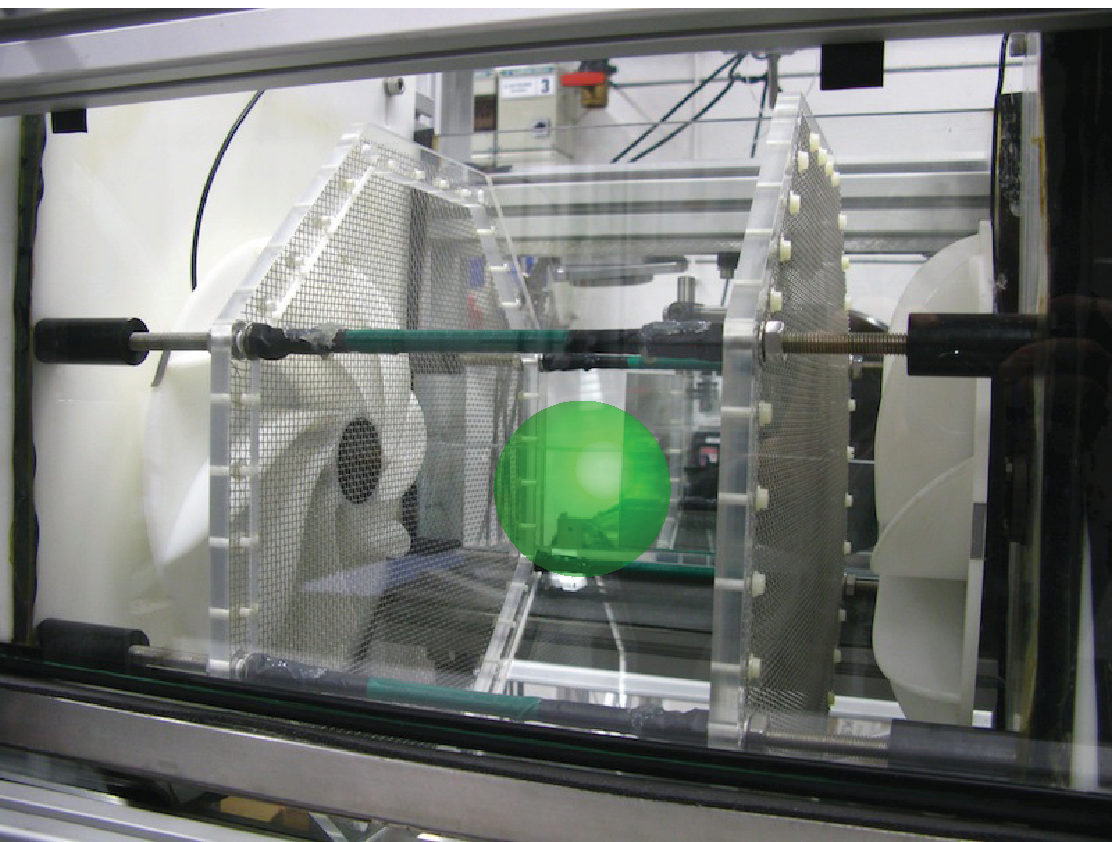}}\quad
  \subfigure[Top view]{\includegraphics[width=0.40\textwidth]{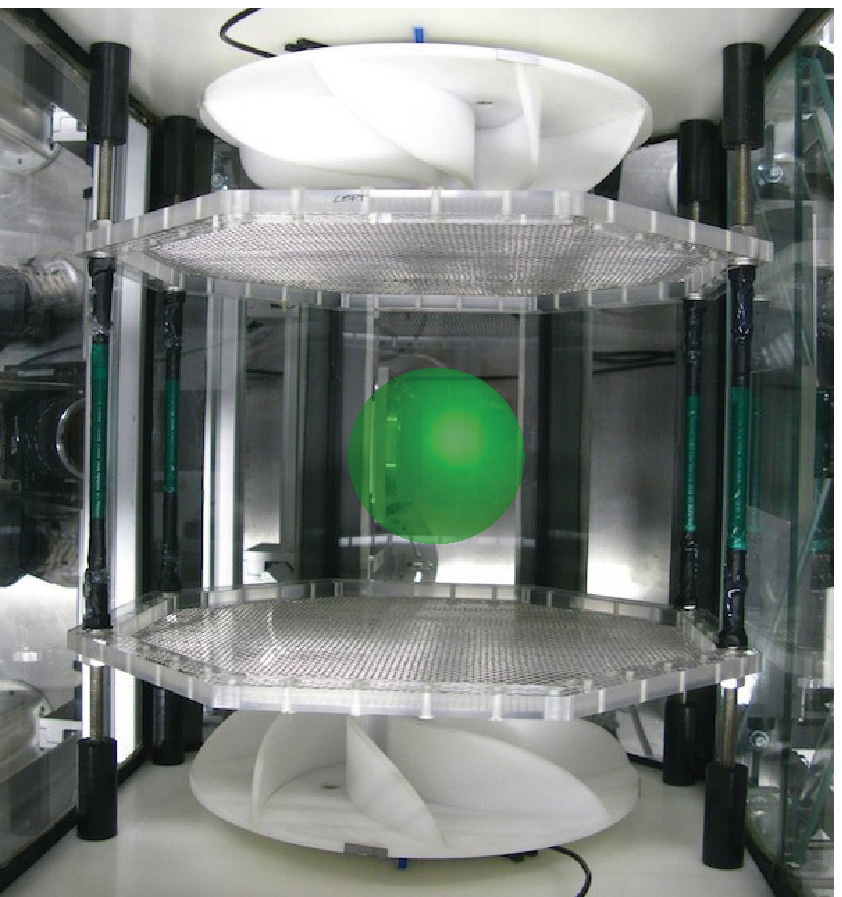}}
\end{center}
\caption{\label{VK_Mixer} Picture of the Von K\'arm\'an water flow. The green sphere at the center of the apparatus represents the measurement volume.}
\end{figure}

In order to generate a highly turbulent flow in a relatively small experiment, we used a von K\'arm\'an water flow, sometimes nick-named the "french washing machine" in the literature (see Fig.~\ref{VK_Mixer})~\cite{Voth:2002p432,Marie_2004,Cortet_2009}. This flow is generated by two counter-rotating propellers of $28~cm$ in diameter facing each other.
The propellers are driven at the same constant rotation rate.
The turbulence chamber, shaped as a octagonal cylinder, measures $40~cm$ along the axis of the propellers and $38~cm$ in both height (vertical) and width (horizontal) in the cross-section.  This apparatus is particularly well suited for LPT studies~\cite{Voth:2002p432} since at the geometric center the mean flow velocity is zero. Therefore, at the center the particle dynamics is mainly driven by the turbulent velocity fluctuations (the large scale flow being deterministic\footnote{The ratio between the average velocity and the velocity fluctuations $\tau = \langle u_{i}\rangle / \sqrt{\langle (u_{i}-\langle u_{i}\rangle)^{2}\rangle}$ does not exceed $25\%$ over the entire measurement volume.})\cite{Marie_2004,Cortet_2009}.\\

To study the turbulent flow field, we seeded the water flow with red fluorescent polymer micro-spheres with a diameter of $107~\mu m$ and a density of $\rho_{tr} = 1.05~g/cm^{3}$\footnote{These fluorescent particles are hard-dyed (internally-dyed) polymer particles which utilise the Firefli process to incorporate the dye throughout the polymer matrix. This method produces bright fluorescent colors, minimises photo-bleaching, and prevents dye leaching into aqueous media. They are made of polystyrene and are sold by \textit{Thermo scientific}. Their absorption (emission) wavelength peaks around $542~nm$ ($612~nm$).}. These particles behave as tracer particles  since  in our flow the Kolmogorov length scale is $\eta\approx 100~\mu m$ (see Tab.~\ref{tab:ExperimentParam}) and the fluid density is $\rho_{f} = 1~g/cm^{3}$. We measured three-dimensional particle trajectories with high spatial and temporal resolutions using LPT~\cite{Ouellette:2006p384,Xu:2008p1921} with three high-speed CMOS cameras (Phantom V10, manufactured by Vision Research Inc., Wayne, USA). The particle velocities and accelerations were then obtained by smoothing and differentiating the trajectories \cite{Mordant:2004p245}. The measurement volume of the LPT was determined as the largest sphere that fits inside the complex volume defined by the intersection of the fields of view from all three cameras and the expanded laser beam. In this experiment, the sphere was $7~cm$ in diameter slightly smaller than the integral length scale $L = (9.3 \pm 0.4) cm $. The frame rate was set to $2.9~kHz$ with a resolution of $768 \times 768$ (each pixel corresponds roughly to $100\mu m$ in space). Given the massive amount of data required to conduct this experiment, we used the weighted averaging algorithm (simple and efficient) to locate the particles onto the 2D images. The number of particles tracked per frame was of the order of $200$. In this configuration, as demonstrated in \cite{Ouellette:2006p384}, the 2D particle finding algorithm and 3D stereoscopic reconstruction processes ensure that the particle positions are recovered with an accuracy of $0.1$ pixels $\approx 10~\mu m$. The relatively low seeding density further guaranteed that, using a "3 frames minimum acceleration" scheme~\cite{Ouellette:2006p384}, we obtained long and high quality particle tracks.

\begin{table}
\begin{center}
\begin{tabular}{c c c c c c}
\hline
\multicolumn{6}{c}{$R_{\lambda}$=$374\pm 8$}\\
\hline
 $u'$ & $\epsilon$ & $L$ & $\eta$ & $\tau_{\eta}$ & $N_{f}$ \\
\small{($m/s$)}&\small{($m^{2}/s^{3}$)}&\small{($mm$)}&\small{($\mu m$)}&\small{($ms$)}&\small{(fr$/\tau_{\eta}$)}\\
\hline
\hline
$0.1$&$0.011\pm 4.10^{-4}$&$93\pm 4$&$98\pm 1$&$9.5\pm 0.2$&$27$\\
\hline
\end{tabular}
\end{center}
\caption{Parameters of the experiment. $u'$ is the root-mean-square of the velocity. $\epsilon$ is the  energy dissipation rate per unit mass. $L\equiv u'^{3}/\epsilon$ is the integral length scale. $\eta\equiv (\nu^{3}/\epsilon)^{1/4}$ and $\tau_{\eta}\equiv (\nu/\epsilon)^{1/2}$ are the Kolmogorov length and time scales, respectively, where $\nu$ is the kinematic viscosity of the fluid. $N_{f}$ is the frame rate of the camera, in frames per $\tau_{\eta}$ and $R_{\lambda} = (15u'L/\nu)^{1/2}$ is the Taylor scale Reynolds number of the flow.}
\label{tab:ExperimentParam}
\end{table}

We measured $u'$ the velocity fluctuation and determined the  energy dissipation rate per unit mass  $\epsilon$ from the inertial range scaling of  the second-order longitudinal and transverse Eulerian velocity structure functions from tracking tracer particles. As explained in details in~\cite{EPL_particle_2010}, we also checked the consistency of the value using two exact inertial range relations: the Kolmogorov's "four-fifth law" and a theorem on the velocity-acceleration mixed structure function~\cite{Mann:1999p3829,Pumir:2001p2611,Hill:2006}. The properties of the turbulent flow field are summarized in Table~\ref{tab:ExperimentParam}.

\begin{figure}
\begin{center}
  \subfigure[]{\includegraphics[width=0.5\textwidth]{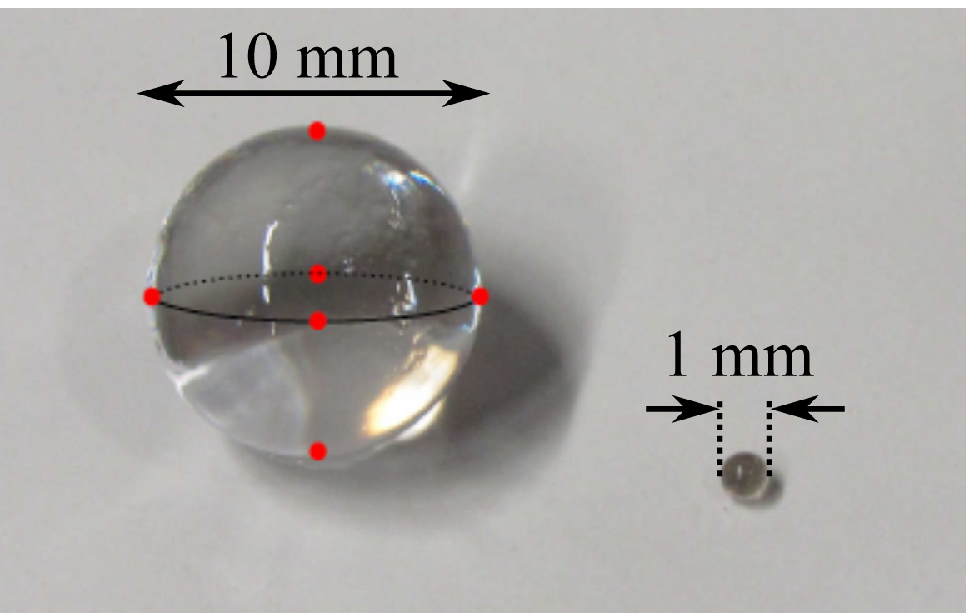}}\quad
  \subfigure[]{\includegraphics[width=0.46\textwidth]{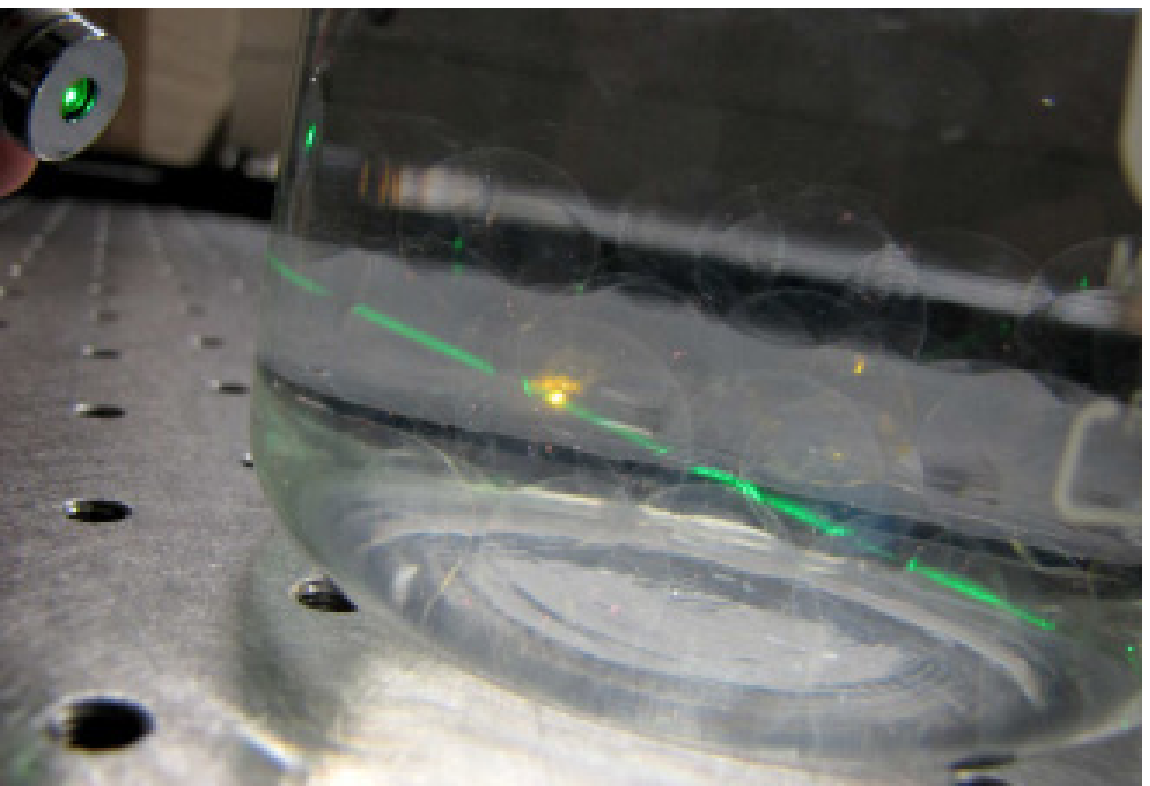}}
\end{center}
\caption{\label{Particles} Super-absorbent polymer particles. (a) Once immersed in water and in their dry form. (b) Beaker full of these particles crossed by a laser beam that is not refracted. The red luminescent dot in the center of the picture is a marker particle grafted at the surface of the solid particle (symbolized by the red dots in (a)).}
\end{figure}

The aim of the investigation was to simultaneously  measure the 3D trajectories of the tracer particles and the full dynamics and trajectories of finite size, inertial particles. The most straightforward way was for the particles  to be invisible in the fluid and to mark them with tracer particles. This avoided  optical distortions and shadowing.  The spherical particles were made of  a super-absorbent polymer (poly-acrylate) from \textit{Aqualinos}, which in their dry form had a diameter between $1$ and $2~mm$; once immersed in water their average diameter grew to $1~cm \approx L/10 \approx 100\eta$ (see Fig.~\ref{Particles}~(a)). Thus the gel particles were 99.9\% water and had almost the same refractive index and density as water.  As shown in Fig.~\ref{Particles}~(b) they are indeed almost invisible in water and do not refract the light path.\\ In order to be able to observe the big particle's motion with the cameras, we grafted their surface with the same  fluorescent particles that we used to track the turbulent flow field. As we will see later, we needed at least $4$ of these markers per inertial particle in order to recover the particle's center and radius. Therefore,  we injected them with $6$ to $8$ markers, as shown by the red dots in Fig.~\ref{Particles}~(a)\footnote{The regular separation presented on Fig.~\ref{Particles}~(a) is not necessary, therefore we do not enforce it.}. To achieve this we employed a lancet needle normally used by diabetics. We dipped it into the tracer particle powder to collect some of them on its tip and penetrated the surface of the big particle. After removing the lancet,  tracer particles were left behind  within the polymer gel. 
We quantitatively checked that the markers were indeed grafted to the big particles by injecting some of them exactly as depicted in figure \ref{Particles} (a) thanks to a specifically designed injection guide. No relative nor absolute displacement could be detected from this well defined injection pattern.
We marked about $150$ big particles, which we inserted into the turbulent flow. This particle number  was necessary to achieve that statistically, at any given time,  at least one (and no more than two) big particles were in the measurement volume.\\

The big solid gel-particles were elastically deformable, but 
the forces (shear and pressure) applied by the fluid onto them were not sufficient to deform them or trigger any internal flows (the internal effective viscosity of the gel being much greater than viscosity of the water). In fact,
from measuring marker positions no deformation could be detected within the experimental uncertainties, while the particles were carried  by the turbulent flow. 
Therefore, these gel-particle can, for the purpose of this study, be considered as solid.
To prevent the particles from being damaged by the propellers, we  placed grids (of $2\times 2~mm^{2}$ mesh size and $75\%$ aperture) in front of both of them (see Fig.~\ref{VK_Mixer}). Measurements showed that the influence on the flow was negligible. We ensured that the measurement volume was sufficiently far from the grid so that the large particles' dynamics was not influenced by the meshes. This can be seen from the following analysis: the viscous relaxation time of the particle is,
\begin{equation}
\tau_{\nu}\equiv\frac{1}{18}\left(\frac{\rho_{p}}{\rho_{f}}\right)\frac{d_{p}^{2}}{\nu}
\end{equation} 
where $\rho_{p}$ and $\rho_{f}$ are respectively the particle and fluid density, $d_{p}$ the particle diameter and $\nu$ the kinematic viscosity of the fluid. This applies only  when the particle Reynolds number ($Re_{p}\equiv u'd_{p}/\nu$) is very small. Following \cite{Clift:1978,Xu:2008p1025}, one can show that the relevant time scale that takes into account small but finite particle Reynolds number can be written as:
\begin{equation}
\tau_{p} = \frac{\tau_{\nu}}{1+0.132Re_{p}^{n}}
\end{equation}
where the exponent $n$ also depends on $Re_{p}$: $n=0.82-0.05\log_{10}Re_{p}$. With $\rho_{p}/\rho_{f}\approx 1$, $d_{p}=0.01~m$, $\nu=10^{-6}m/s^{2}$, and $u'=0.1~m/s$ this time scale is $\tau_{p}=0.38~s$, which yields $L_{p}=u'\tau_{p}=3.8~cm$ beyond which the big, inertial particle has reached the fluid velocity. The minimal distance between the grid and the measurement volume is $\approx 7~cm$ on both sides and thus we expect that the particles' dynamic is not impacted by the screens.\\

As pointed out before, we were seeding the turbulent flow with two types of particles: big particles marked by fluorescent tracers and the  fluorescent tracers themselves. When tracking all fluorescent particle tracks where visible simultaneously; marker particles trapped in the big particles did not change their separation, while  tracer particles  would separate quickly.

\section{Identifying the finite-size particles}
\label{identifyProcess}
One very significant advantage of the novel technique presented here is that it allows to follow two species of particles using only one LPT system. For comparison, Guala et al. ~\cite{Guala:2008p2698}  used two LPT systems, which makes such an experiment very demanding. 
To identify the two populations of particles we took advantage of the fact that the flow was highly turbulent. It is well known that tracer particles separate very quickly \cite{Bourgoin:2006p391,EPL_particle_2010}.  In the contrary markers fixed to the surface of a big particle will stay at a constant distance. Specifically, two tracer particles initially separated by a distance $R_{0}\equiv |\mathbf{R}(t=0)|$ in the inertial range (\textit{i.e.} $\eta\ll R_{0}\ll L$) will separate on average following a ballistic regime predicted by Batchelor \cite{batchelor:1950} such as:
\begin{equation}
\langle\delta\mathbf{R}\cdot\delta\mathbf{R}\rangle = \frac{11C_{2}}{3}R_{0}^{2}\left(\frac{t}{t_{0}}\right)^{2}
\label{Batchelor}
\end{equation}
for $t\ll t_{0} = (R_{0}^{2}/\epsilon)^{1/3}$, where $\delta\mathbf{R}(t)\equiv \mathbf{R}(t)-\mathbf{R}(t=0)$ is the vectorial separation increment and $C_{2}=2.1$ as suggested from a compilation of available data \cite{sreenivasan:1995}.\\

\subsection{Finding pairs of tracks}
\label{Pairs}
The first step is to find pairs of fluorescent particles tracks whose distance to each other stays constant during the time of observation. As the fluorescent markers are embedded in the particles with radius $R_p$, we scan only for tracks whose separations are  $0.05<R_{t}/R_{p}<2$(here $R_{p}=8~mm$). The lower bound comes from the fluorescent particles injection process that we used. Additionally, to limit possible false choices, we only conduct this analysis on pairs of tracks that coexist for longer than $2\tau_{\eta}$. This is the characteristic time-scale after which the acceleration components of a fluid particle decorrelated entirely~\cite{Voth:2002p432}. To estimate the tolerance on the separation increments, we have used Equ.~\ref{Batchelor} with $R_{0}=0.05R_{p}$, $t=2\tau_{\eta}$ and $\epsilon=0.011~m^{2}/s^{3}$ which gave us $\sqrt{\langle\delta\mathbf{R}^{2}\rangle}\approx 0.1R_{p}$. In the analysis presented here we used a slightly tighter criterion: $\sqrt{\langle\delta\mathbf{R}^{2}\rangle} < 0.01R_{p}$, that corresponds approximately to our spatial resolution ($100~\mu m/pixels$). 

\subsection{Finding groups}
\label{findgroup}
To ease the explanations, two tracks that were paired by the algorithm described  above (section~\ref{Pairs}) are now called "friends". We can now identify the fluorescent particle tracks that belong to a unique inertial particle. To do so we simply apply the adage "the friend of a friend is my friend". This allows to generate groups of tracks that trace the inertial  particles'  motion. By additionally imposing that at least four tracks  coexist for more than $2\tau_{\eta}$ we exclude optical artifacts. Such a group is presented on Fig.~\ref{Reconnect} (b).

\subsection{Track reconnection inside the groups}
\label{recon}
\begin{figure}[h!]
\begin{center}
  \subfigure[]{\includegraphics[width=0.48\textwidth]{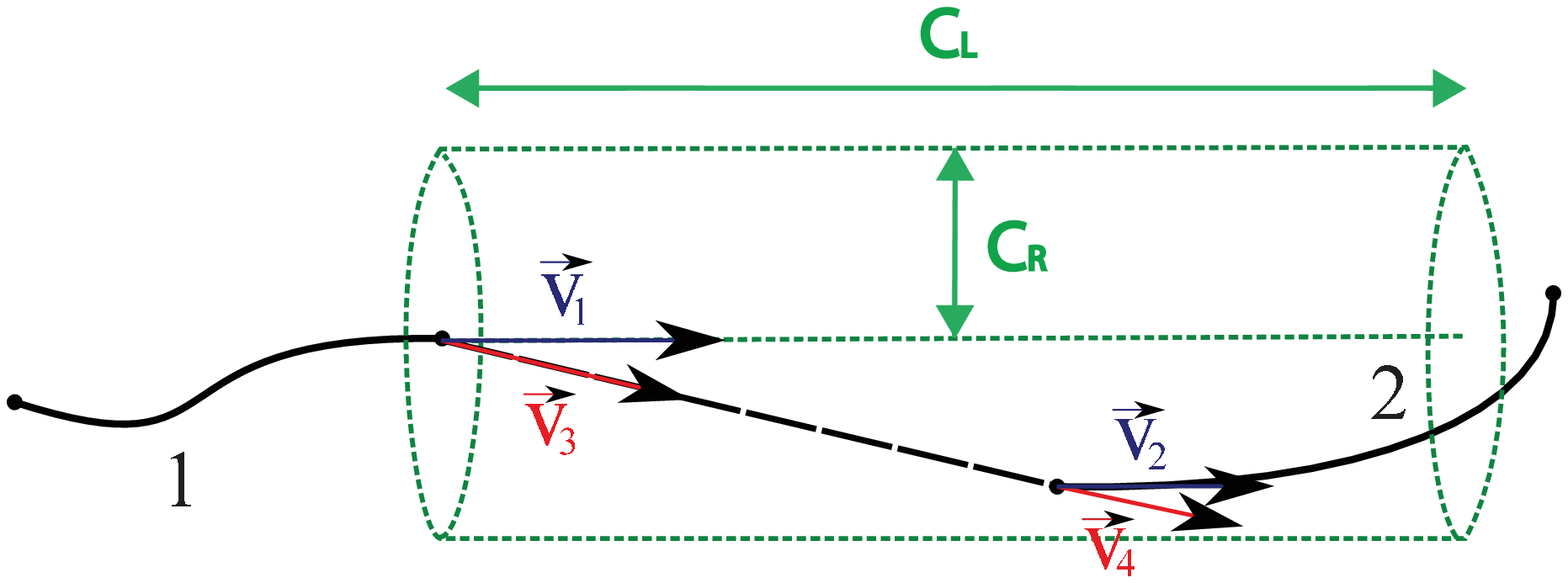}}\quad
  \subfigure[]{\includegraphics[width=0.23\textwidth]{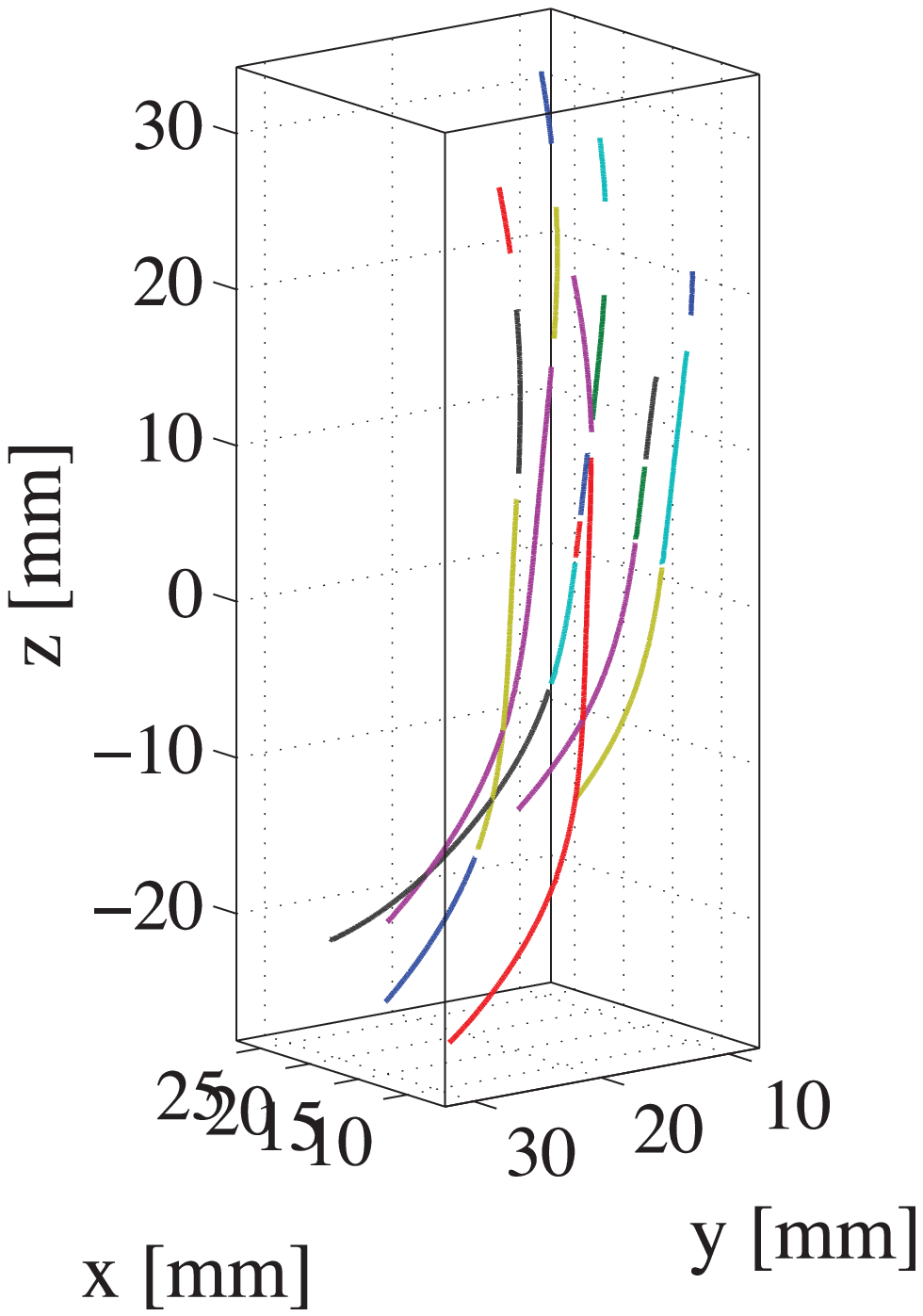}}\quad
  \subfigure[]{\includegraphics[width=0.23\textwidth]{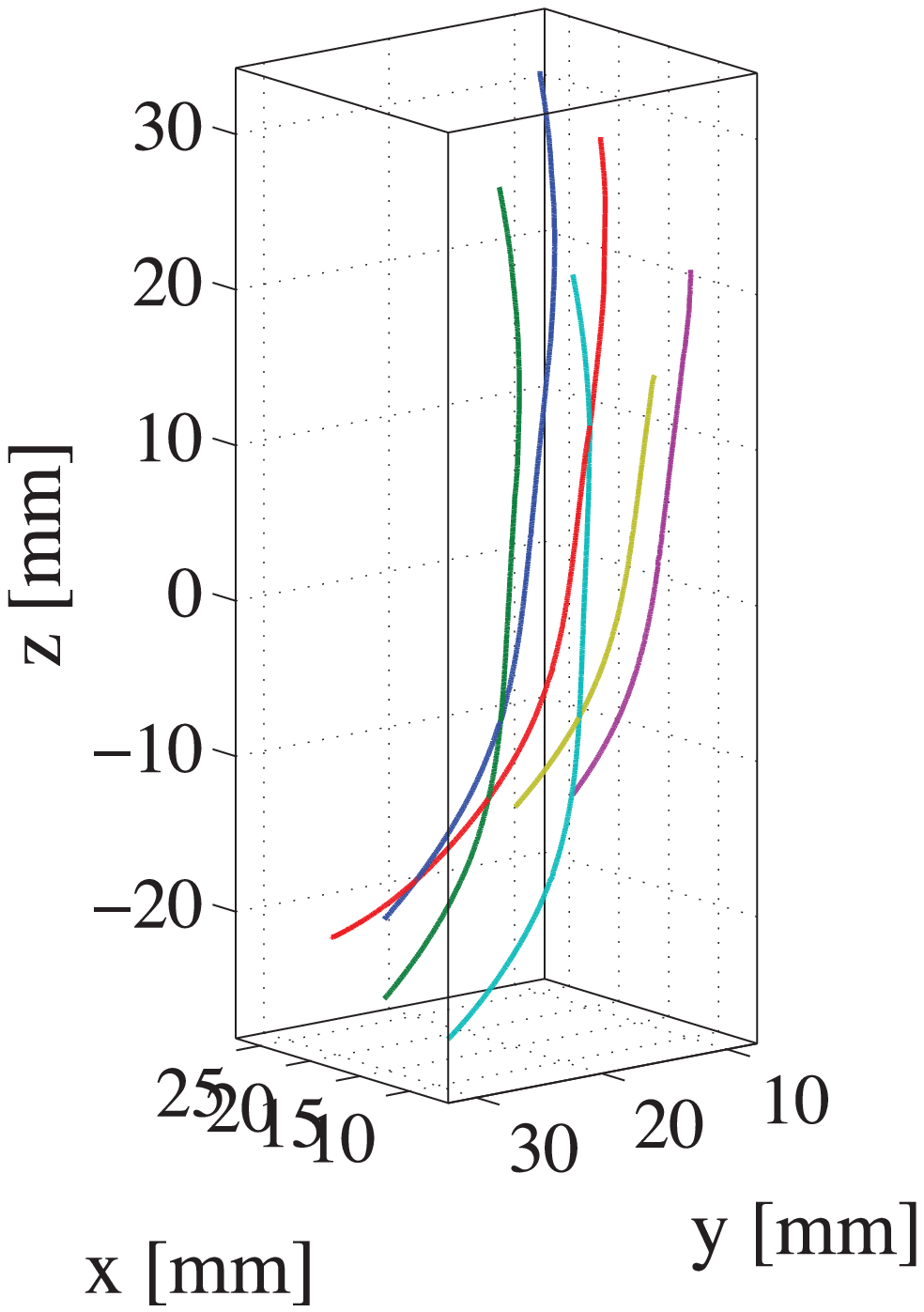}}\\
  \subfigure[]{\includegraphics[width=0.48\textwidth]{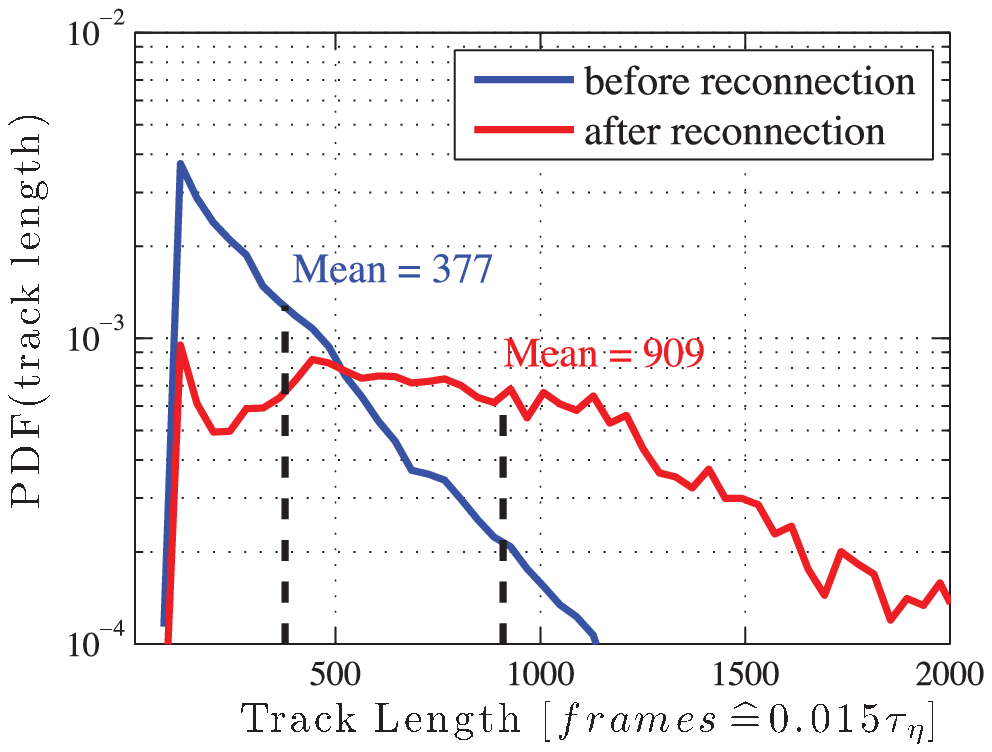}}
\end{center}
\caption{\label{Reconnect} Track reconnection inside the groups. (a) Definition of the search volume for possible track reconnection. (b) Group before reconnection. (c) Group after reconnection. (d) Probability density function of the track length within the groups, before and after reconnection.}
\end{figure}

As one can see from Fig.~\ref{Reconnect} (b), the tracks that belong to a group occur in short segments. This can be attributed to  limitations of the experiment, which include variations of the illumination intensity, particles blocking the line of sight of the cameras,  particles blocking the illumination beam (shadowing), the presence of light-insensitive circuits on the CMOS sensors of the cameras, and the effect of thermal, electronic and environmental noise (see~\cite{Xu:2008p1921}). To  reconnect the tracks we use an algorithm similar to that  developed by  Haitao  Xu at the Max Planck Institute for Dynamics and Self-Organization~\cite{Xu:2008p1921}. Within a group, at the end of each track, we define a cylindrical search volume, parametrized by its length $C_{L}$, its radius $C_{R}$ and its axis oriented along $\mathbf{V_{1}}$ (the velocity at the end of the track-segment). In Fig.~\ref{Reconnect}~(a) such a cylinder is shown exemplarily in green. Tracks of the group that have their starting point within this search volume are considered potential re-connection candidates (track $2$ on Fig.~\ref{Reconnect}~(a)). Then for each candidate, we compute the linear interpolation of the trajectory and compare the beginning and ending interpolated velocities (resp. $\mathbf{V_{3}}$ and $\mathbf{V_{4}}$ on Fig.~\ref{Reconnect}~(a)) to the actual measured ones (resp. $\mathbf{V_{1}}$ and $\mathbf{V_{2}}$ on Fig.~\ref{Reconnect}~(a)). If the beginning and ending relative velocity differences are smaller than a tolerance $C_{T}$ (\textit{e.g.} $|\mathbf{V_{3}}-\mathbf{V_{1}}|/|\mathbf{V_{1}}|<C_{T}$) we consider these as a possible reconnection. If there are several candidates, we re-connect the best match. We remind the reader that we are working within a group obtained by the previous step (section \ref{findgroup}), that is to say a very small amount of tracks compared to the work in~\cite{Xu:2008p1921}.  In our experiments, we found that  $C_{R}=0.5~mm$, $C_{L}=30~mm$ and $C_{T}=0.5$ give satisfactory results. For example, we present the result of this reconnection process on a particular group in Fig.~\ref{Reconnect}~(c), and its statistical effect on several measurements in Fig.~\ref{Reconnect}~(d) (please note that this more than doubles  average track length inside the groups).

\subsection{Cleaning the groups}
\label{cleaning}
The process described above is not error-proof. Indeed, while scanning examples of groups by eye we could identify wrong tracks. We therefore developed the "group cleaning" technique. In this step, we scan each group and we keep only tracks that have more than three direct "friends" as defined in section~\ref{Pairs}. We use the same parameters except that we relax the constraint on the maximal separation increment to $\sqrt{\langle\delta\mathbf{R}^{2}\rangle} < 0.07R_{p}$.  This we can do as the data was already connected by linear interpolation. 

\subsection{Concluding remarks on the particle tracking procedure }
\label{ClPro}
\begin{figure}[h!]
\begin{center}
  \subfigure[]{\includegraphics[width=0.48\textwidth]{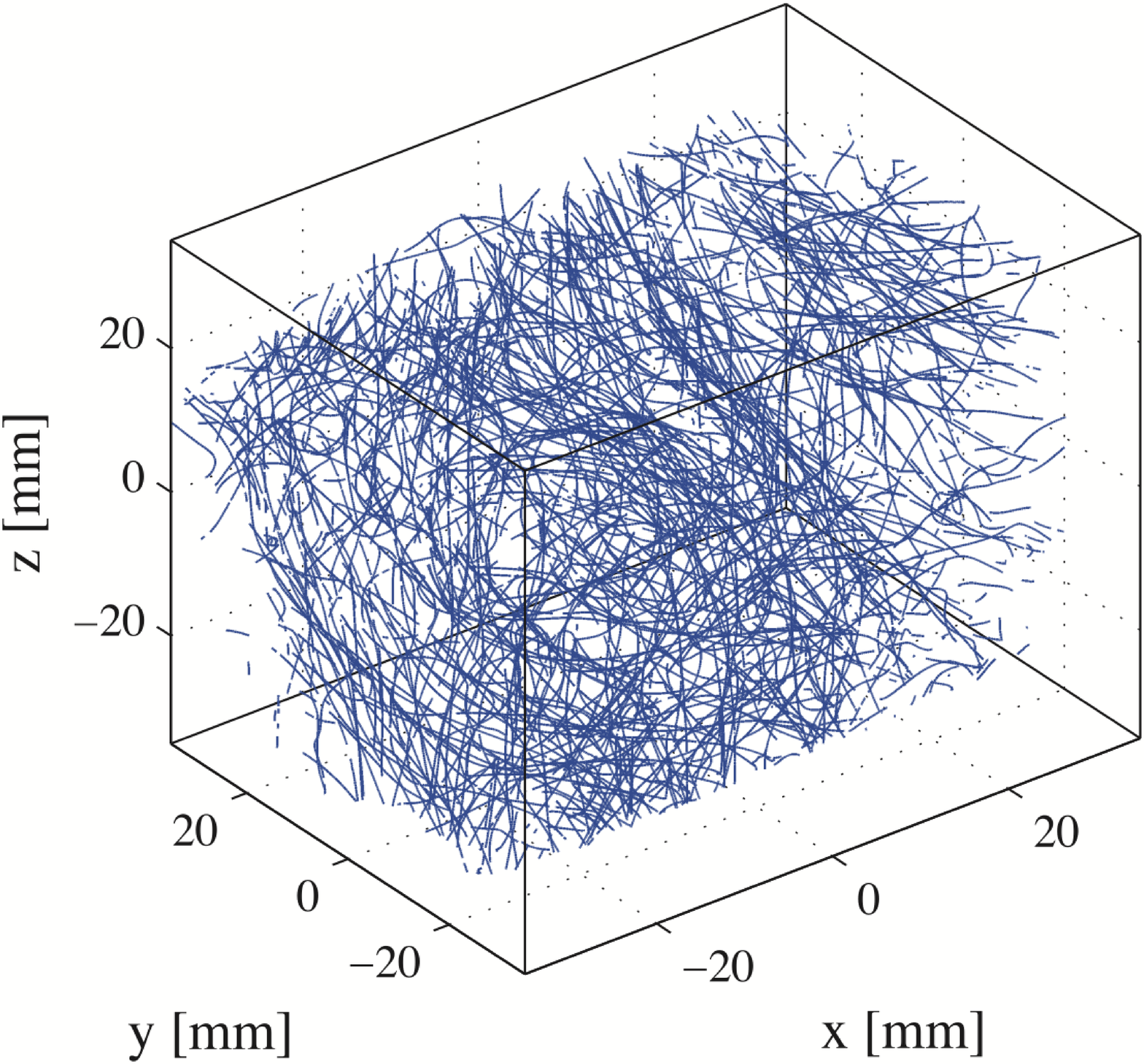}}\quad
  \subfigure[]{\includegraphics[width=0.48\textwidth]{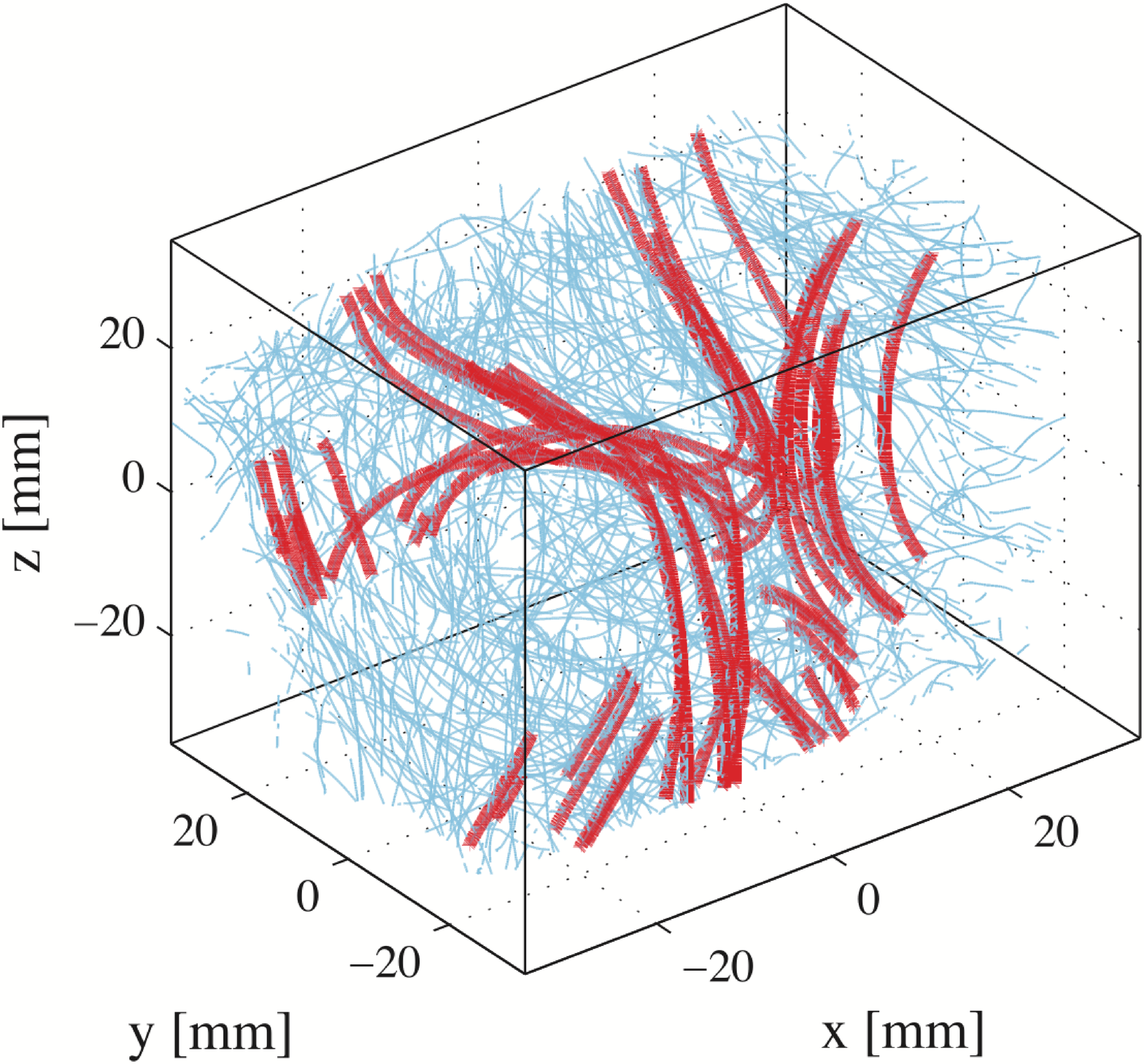}}\quad
  \end{center}
\caption{\label{identify} Result of the process described in section \ref{identifyProcess} on a $3~s$ movie. (a) All the trajectories of the fluorescent particles; (b) Same data set in light blue. The red trajectories correspond to the fluorescent particles bounded to the surface of the super-absorbent polymer spheres identified by the procedure described in section~\ref{identifyProcess}.}
\end{figure}

As shown by Fig.~\ref{identify} we can  now separate the two populations of fluorescent particles. In this process, as discussed above,  we  interpolate  trajectories (section~\ref{recon}) solely to connect tracks.  These interpolated  data points need to be excluded when statistics of the fluid velocity, or vorticity along tracks is considered.  We would also like to note that an inspection by eye of  $\approx 5000$ groups did show the  procedure described above works only well, when only one inertial particle is in the measurement volume at any given time. We therefore  discarded manually the rare cases where two or more particles (some colliding) were in view.  This did not limit the statistics significantly. In the future it will be not hard to modify the algorithm to include this automatically.  

\section{Translation of the finite size particle}
\label{translation}
Here, we describe how we extract the center position $\mathbf{x_{c}}$ and radius $R_{p}$ of the big inertial particles for each group identified by the procedure described in section~\ref{identifyProcess}. These groups have a minimum of four marker-trajectories. As described above,  our injection technique ensured reproducible and well defined penetration depth into the inertial particle. 
The injection was such that the markers were inserted not more than  $0.5~mm$ from the polymer surface. The center position $\mathbf{x_{c}}$ and radius $R_{p}$ of the inertial particle is entirely defined by the sphere equation:
\begin{equation}
(x_{i}-x_{c})^{2}+(y_{i}-y_{c})^{2}+(z_{i}-z_{c})^{2}=R_{p}^{2}
\end{equation}
where $(x_{i}; y_{i}; z_{i})$ are the $i$-th marker's  3D-coordinates. For each permutation of four of these markers, one gets the following linear system:
\begin{equation}
\underline{\underline{\bold{P}}}\cdot\left(A~~~B~~~C~~~D\right)^T=\text{Tr}(\underline{\underline{\bold{X}}}\cdot\underline{\underline{\bold{X}}}^T)
\label{linearsyst}
\end{equation} 
with
\begin{equation}
\underline{\underline{\bold{P}}}\equiv\left(\begin{array}{c c c c}-1~~x_1~~ y_1~~ z_1\\-1 ~~x_2~~ y_2 ~~z_2\\-1~~ x_3 ~~y_3~~ z_3\\-1 ~~x_4 ~~y_4~~ z_4\end{array}\right)\hspace{3cm}\underline{\underline{\bold{X}}}\equiv\left(\begin{array}{c c c}x_1~~ y_1~~ z_1\\x_2~~ y_2 ~~z_2\\x_3 ~~y_3~~ z_3\\x_4 ~~y_4~~ z_4\end{array}\right)
\end{equation}
and $A\equiv x_{c}^{2}+y_{c}^{2}+z_{c}^{2}-R^{2}$, $B\equiv -2x_{c}$, $C\equiv -2y_{c}$ and $D\equiv -2z_{c}$. Let us consider a group that contains $n\geq 4$ markers trajectories at a given time $t$, the number of permutations of $4$ trajectories out of $n$ is $M=\frac{n!}{4!(n-4)!}={n \choose 4}$. Therefore, at each time step, we solve $M$ times the linear system given by Equ. \ref{linearsyst}. We define $\mathbf{x_{c,m}}(t)$ the center coordinates obtained by the $m$-th permutation and $d_{m}(t) = \sum_{i=1}^{4}\sum_{j=i+1}^{4}|\mathbf{x_{i,m}}(t)-\mathbf{x_{j,m}}(t)|$ the sum of the distances between the markers. To define the actual center coordinates of the inertial particles, we perform an average of $\mathbf{x_{c,m}}(t)$ weighted by $d_{m}(t)$:
\begin{equation}
\mathbf{x_{c}}(t)=\sum_{m=1}^{M}\mathbf{x_{c,m}}(t)d_{m}(t)/\sum_{m=1}^{M}d_{m}(t).
\end{equation}
We have made this choice as to correct for a non-homogeneous distribution of the markers on the inertial particle's surface. 
Indeed, four markers homogeneously distributed around the sphere define a center position and a radius more accurately than four agglomerated points.

\begin{figure}[h!]
\begin{center}
  \subfigure[]{\includegraphics[width=0.48\textwidth]{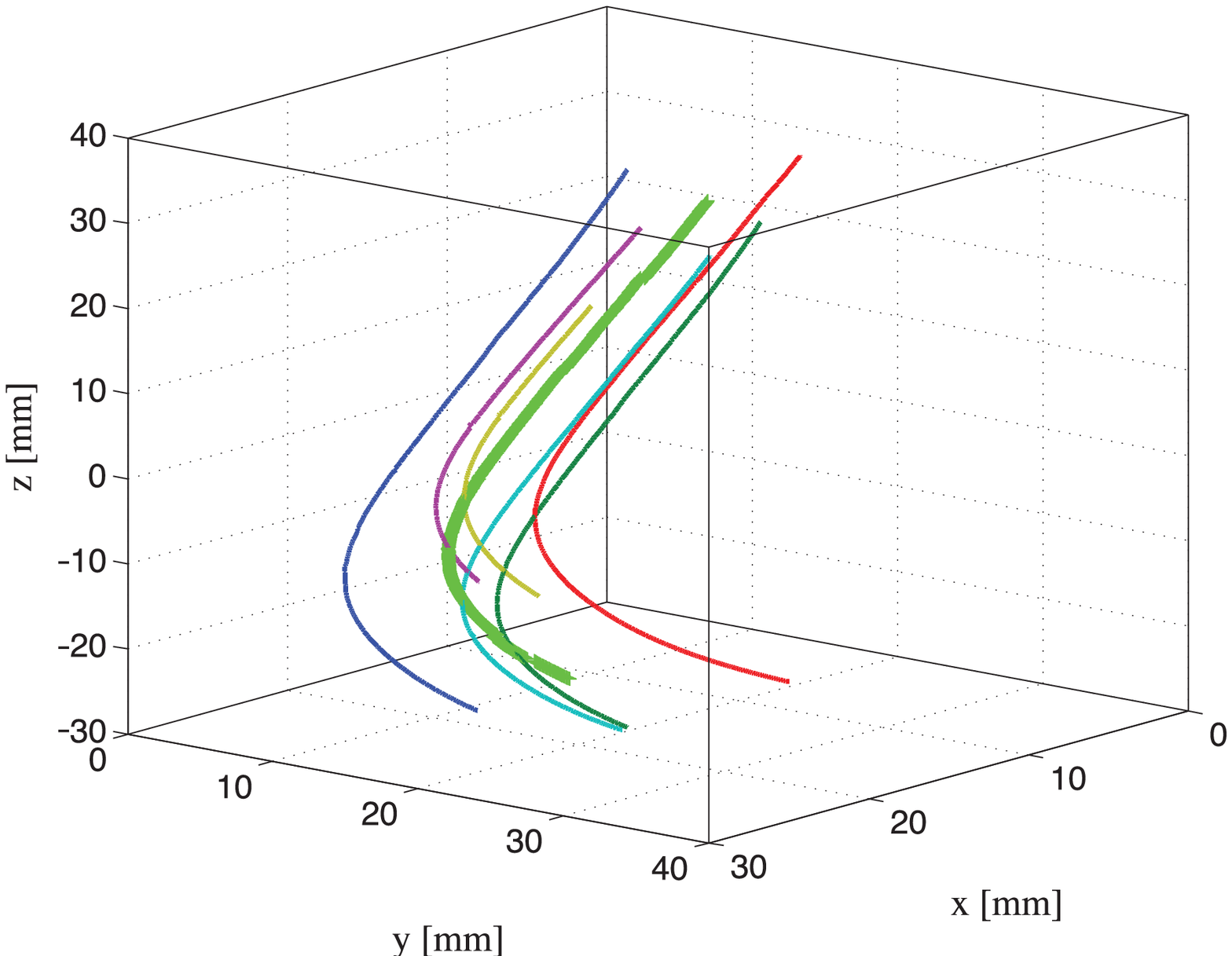}}\quad
  \subfigure[]{\includegraphics[width=0.48\textwidth]{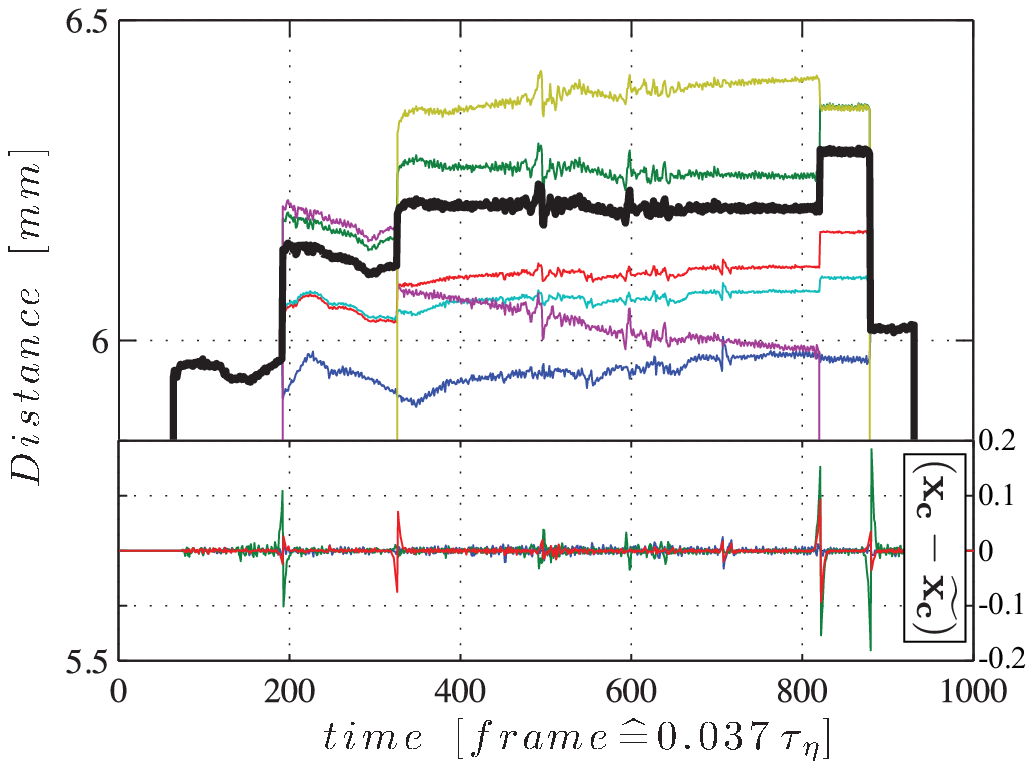}}\quad
  \end{center}
\caption{Results of the center finding process for one example group. (a) Trajectories (thin lines) identified to belong to the particle and the center (thick green line). (b) Radius (bold black line) and the distances from the center to each trajectory (solid lines) for the tracking time of the particle. In this specific case, the particle was tracked for $1115\,\text{ frames} \mathrel{{=}} 41.3\,\tau_\eta\approx 0.4\text{s})$. The inset at the bottom shows the 3 coordinates of particle center $\mathbf{x_{c}}$ subtracted by its filtered version (Gaussian kernel of $\sigma = 3.5 $ frames) $\mathbf{\widetilde{x_{c}}}$, in [$mm$].}
\label{CenterFinding}
\end{figure}

The result of the center finding process on a particular group is shown on Fig.~\ref{CenterFinding}~(a), together with the radius of the inertial  particle  $R(t)$, and the distances from the center position to the markers of this group $r_{i}(t)=|\mathbf{x_{i}}(t)-\mathbf{x_{c}}(t)|$ in Fig.~\ref{CenterFinding}~(b). The latter shows that we are able to measure the particle radius with an uncertainty of $\pm 200\mu\text{m}$, which corresponds to the uncertainty of the fluorescent marker insertion. We also observed that  $R(t)$ exhibits unrealistic rapid variations when a marker track appear or disappear from the group (around frames $200$, $350$, $800$ and $900$). These kinks are non-physical and result from the uncertainty of the marker position inside the inertial particle.\\
Just as for tracer particles  \cite{Mordant:2004p245},  we measured the velocity  of inertial particles from the center trajectory by convolving with a properly normalized, truncated, differentiating Gaussian smoothing kernel. The characteristic time scale of the filter was chosen as $\sigma=3.5 \text{ frames} = 0.13\tau_{\eta}$ sufficient to minimize the effect of noise on the tracer particles' acceleration variance. At the kinks observed in the center position of the big particles $\mathbf{x_{c}}(t)$ (Fig.~\ref{CenterFinding}~(b)) the values of the  velocities were very large, discontinuous, and thus unphysical. We flagged and excluded those from the analysis 
together with $2\sigma = 7$ frames before and after those discontinuities.

\section{Rotation of the inertial particle}
\label{rotation}

To measure the rotation of the inertial particles, we used the center position  together with the trajectories of the markers around the center of the sphere. To do so, we took advantage of an algorithm first introduced by  W. Kabsch~\cite{Kabsch1976,Kabsch1978} that was developed to compare molecular conformations in chemistry. This algorithm seeks the optimal rotation matrix $U$ between two sets of points by minimizing the root mean square of their separation.\\
Let us consider two sets $x$ and $y$ of $N$ paired points that have the same centroid at the origin. We are looking for the $3\times 3$ unitary matrix $U$ that aligns best $x$ with $y$. This can be achieved by minimizing the root mean squared:
\begin{equation}
D=\left[\frac{1}{N}\sum_{i=1}^{N}(Ux_{i}-y_{i})^{2}\right]^{1/2}
\end{equation}
That can be rewritten as:
\begin{equation}
ND^{2}=\sum_{i=1}^{N}[(Ux_{i})^{2}+y_{i}^{2}-2Ux_{i}y_{i}]
\end{equation}
where one can see that minimizing the left hand side is equivalent to maximize the negative term on the right hand side. If we represent the points of set $x$ (resp. $y$) by a $N\times 3$ matrix $X$ (resp. $Y$), the quantity to maximize is:
\begin{equation}
\sum_{i=1}^{N}Ux_{i}y_{i} = \text{Tr}(Y^{T}UX) = \text{Tr}((XY^{T})U)
\label{Maximise2}
\end{equation}
where $XY^{T}$ is a square $3\times 3$ matrix, that can be rewritten using its singular value decomposition (SVD) as $XY^{T}=VSW^{T}$. $V$ and $W^{T}$ are orthogonal matrices of the left and right eigenvectors of $XY^{T}$ and $S$ is a diagonal $3\times 3$ matrix containing the eigenvalues $s_{1}\geq s_{2}\geq s_{3}$. Using the commutation properties of the trace operator, Equ.~\ref{Maximise2} can be rewritten as:
\begin{equation}
\text{Tr}(VSW^{T}U) = \text{Tr}(SW^{T}UV) = \sum_{i=1}^{3}s_{i}T_{ii}
\label{Maximise3}
\end{equation}
where $T=W^{T}UV$ is a product of orthogonal matrices and itself an orthogonal matrix with $\text{det}(T)=\pm 1$. Therefore each elements of $T$ are equal or smaller than $1$. Thus to maximize Eq.~\ref{Maximise3}, $T$ has to be the identity matrix $T=I$.

The last constraint that we have to consider is that 
\begin{equation}
U=WTV^{T}
\end{equation}
has to be a proper rotation matrix with $\text{det}(U)=1$ (rows/columns of $U$ have to form a right handed system)~\cite{Kavraki2007}. For $\text{det}(XY^{T})>0$, it follows that $\text{det}(U)=1$, but when $\text{det}(XY^{T})<0$, $\text{det}(U)=-1$. In the later case, one has to settle for the second largest value of Eq.~\ref{Maximise3}. This value is obtained when $T_{33}=-1$ since $s_{1}\geq s_{2}\geq s_{3}$. This finally allows us to write the optimal rotation matrix $U$ as
\begin{equation}
U=W \left( \begin{array}{ccc}
1 & 0 & 0 \\ 0 & 1 & 0 \\ 0 & 0 & d\\
\end{array}\right) V^T,
\label{OptRot}
\end{equation}
where $d=\text{sign}(\text{det}(XY^{T}))$.\\

When applying  this algorithm to our data, we constructed the $N\times 3$ matrices $X_{ni}=x_{ni}(t)-x_{c,i}(t)$ and $Y_{ni}=x_{ni}(t+\Delta t)-x_{c,i}(t+\Delta t)$, where $x_{ni}$ are the position components of the $n$-th marker trajectory that existed at times $t$ and $t+\Delta t$ ($\Delta t$ is equal to the  acquisition time). By construction  for each  group $N \ge 4$ (see section~\ref{identifyProcess}) thus ensuring a proper rotation matrix. Then for each time steps, we computed the covariance matrix $C=XY^{T}$, its SVD $C=VSW^{T}$ and the sign of its determinant $d=\text{sign}(\text{det}(C))$. Next, we applied  Equ.~\ref{OptRot} to obtain the optimal rotation matrix $U$. 

\begin{figure}[h!]
\begin{center}
\subfigure[]{\includegraphics[width=0.48\textwidth]{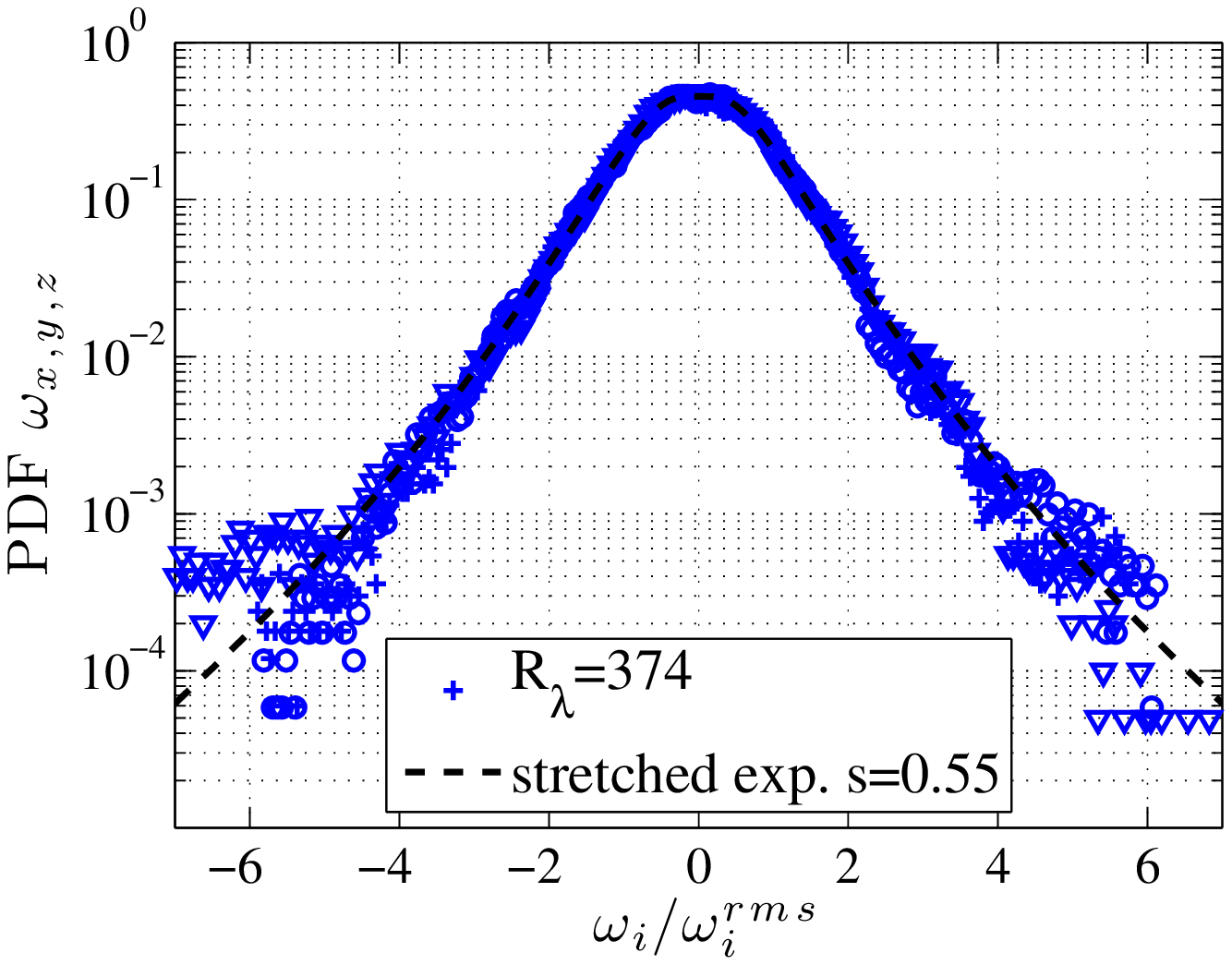}}\quad
  \subfigure[]{\includegraphics[width=0.48\textwidth]{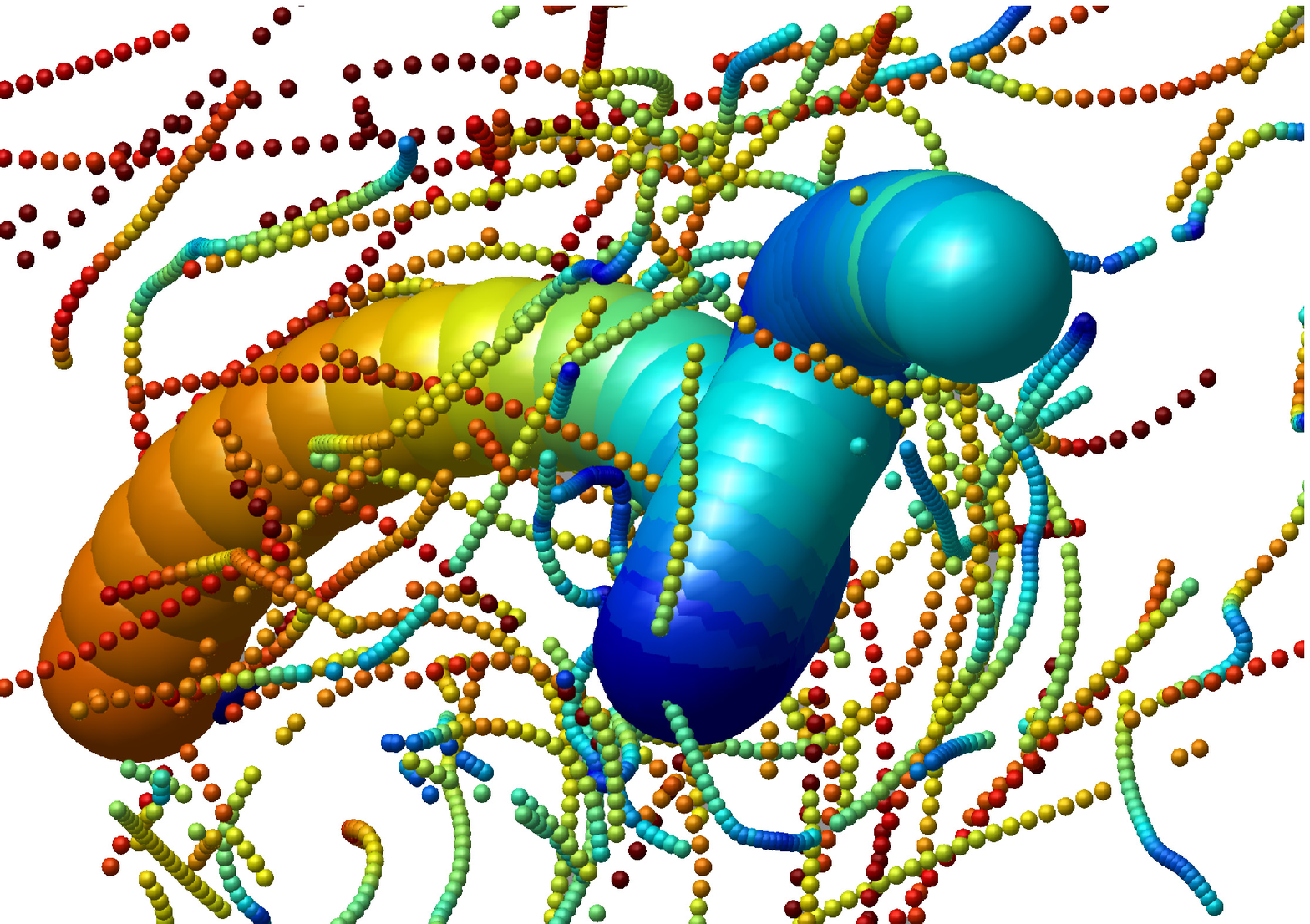}}\quad
  \end{center}
\caption{(a) Reduced probability density function of the rotation rate component of the inertial particle (the different symbols corresponds to the 3 components $\omega_x$, $\omega_y$, $\omega_z$). The dashed line corresponds to a stretched exponential function. (b) 3D visualization of the inertial particle trajectory together with the tracer particles around it. The colors encode the velocities of the particles, warm colors indicate fast particles and cold ones low velocity particles.}
\label{FigRot}
\end{figure}

From the rotation matrix, one can extract the rotation vector or the Euler angles and compute the rotation rate of the inertial particle $\mathbf{\omega}$. 
One can estimate an uncertainty of about $5\%$ on the components of $\mathbf{\omega}$ from the analysis of the residuals of the least square optimisation described above.
Figure~\ref{FigRot}~(a) shows the measured probability density function of the components for the rotation of the inertial particle. This distribution is non Gaussian and can be described by a stretched exponential function $P(x) = \frac{\exp(3s^{2}/2)}{4\sqrt{3}}\left[1-\text{erf}\left(\frac{\ln(|x/\sqrt{3}|)+2s^{2}}{\sqrt{2}s}\right)\right]$ with a best fit parameter $s=0.55$. Zimmermann \textit{et al.}~\cite{2011PhRvL.106o4501Z} used the same functional form with $s=0.45$ which is in good agreement with our results.\\

As shown in Fig.~\ref{FigRot}~(b), we measured simultaneously the full motion of a finite size particle, translation and rotation, and the highly turbulent flow field from tracer particle tracking.  This method  thus allows us to  study experimentally the interactions between the particle and the flow.

\section{Particle flow correlation}
\label{PFcorrel}

This novel measurement technique is particularly well suited to study particle fluid correlation. The quantity that we want to study here is the second moment of the longitudinal velocity difference between the inertial particle and the flow, also called the mixed (particle/flow) Eulerian second order longitudinal velocity structure function:
\begin{equation}
D_{LL}^{mix}(\mathbf{r}) = \langle [(\mathbf{v_{p}}(0)-\mathbf{v_{f}}(\mathbf{r}))\cdot\mathbf{\hat{r}}]^{2}\rangle
\end{equation}
where $\mathbf{v_{p}}$ and $\mathbf{v_{f}}$ are the inertial particle and fluid velocity vectors, and $\mathbf{r}$ is the separation vector between the center of the inertial particle and the tracer particle considered ($\mathbf{\hat{r}}=\mathbf{r}/|\mathbf{r}|$). In general, this quantity depends on the position of the particle and on the separation vector $\mathbf{r}$. Here however,  we measured at  the center of a von K\'arm\'an mixer,  where the flow is fairly homogeneous and isotropic, therefore one can assume that $D_{LL}^{mix}$ is only a function of $\mathbf{r}$. Additionally, given the symmetries of the system, we averaged the data azimuthally around $\mathbf{v_{p}}$ axis, where we expect rotational invariance.

\begin{figure}[t!]
\begin{center}
\includegraphics[width=0.48\textwidth]{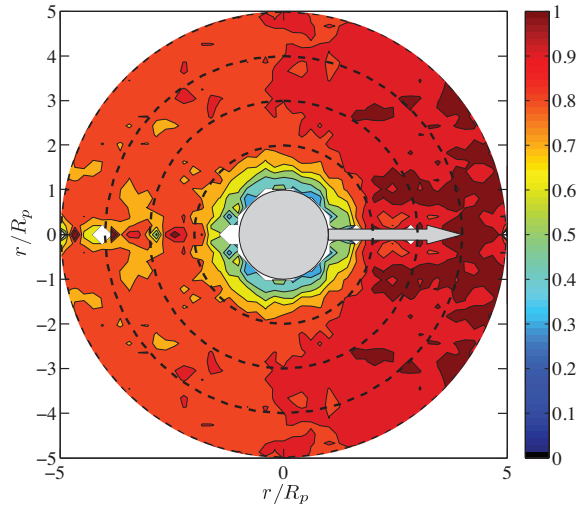}
  \end{center}
\caption{Normalized mixed longitudinal 2nd-order Eulerian velocity structure function $\hat{D}^{mix}_{LL}$. $\hat{D}^{mix}_{LL}$ as a function of the relative separation distance $r/R_p$ (where $R_p$ is the particle's radius) and $\theta$ being the angle between the separation vector $\mathbf{r}$ and the velocity vector of the ball $\mathbf{v_p}$. The orientation of $\mathbf{v}_p$ is shown by the grey arrow (with an arbitrary length). The grey circle in the middle of the map symbolizes the inertial particle. Assuming axis-symmetry around the velocity axis of the particle, the lower half is a reproduction of the upper half mirrored on the center line. The black dashed lines show the separations $r=2R_p,\,3R_p,\,4R_p,5R_p$.}
\label{DLL_2D}
\end{figure}

In order to quantify the influence of the inertial particle on the turbulent flow field, we have normalized $D_{LL}^{mix}$ by its expected value $D_{LL}^{flow}$ in the case of a point-like particle (a tracer/fluid particle since $\rho_{p}/\rho_{f}\approx 1$):
\begin{equation}
\hat{D}_{LL}^{mix}=D_{LL}^{mix}/D_{LL}^{flow}=D_{LL}^{mix}/C_{2}(\epsilon r)^{2/3}
\end{equation}
where $D_{LL}^{flow}=C_{2}(\epsilon r)^{2/3}$ is the second order longitudinal velocity structure function of a homogeneous and isotropic turbulent flow (neglecting intermittency corrections) for separations $r$ in the inertial range ($\eta\ll r\ll L$).\\

$\hat{D}_{LL}^{mix}$ is shown in Fig.~\ref{DLL_2D}. 
The statistical convergence of this quantity, obtained by binning the space around the big particle, was checked by splitting the dataset in two sub-samples, leading to the same conclusions detailed hereafter.
First, one can see that both limits $r=R_{p}$ and $r\longrightarrow\infty$ are consistent with what we would expect. Indeed, $\hat{D}_{LL}^{mix}(r=R_{p})=0$ simply means that no fluid is entering or exiting the solid particle, and $\hat{D}_{LL}^{mix}(r\longrightarrow\infty)=1$ shows that the particle/flow interaction occurs locally. We can identify a spherical shell of radius  $1<r/R_{p}<3$, where the correlation between the inertial particle and the flow is greatly enhanced. The numerical results by  Naso et al.~\cite{Naso:2010p4267} are in good agreement with our experimental observations. Furthermore, beyond an isotropic spherical shell of distance $R_{p}$ from the surface of the inertial particle we observed a  weak anisotropy with respect to the particle direction. This could indicate the presence of a wake induced by the particle.

\section{Conclusion and perspectives}
The main advantages of this new measurement technique are:
\begin{enumerate} 
\item{it requires nothing more than the usual equipment used for "standard" LPT or Tomo-PIV (3 or 4 fast cameras and a light source);} 
\item{it allows to follow in three dimensions the full dynamics (translation and rotation) of a solid body (here a spherical particle) together with the flow field carrying it;}
\item{the solid object, being made out of super-absorbent polymer, has the same index of refraction as the fluid which makes it invisible and therefore it does not block the fields of view of the cameras (no shadowing effects);} 
\item{the geometry of the solid object is recovered during the acquisition (in this case the particle radius $R_{p}$) and could be studied dynamically;}
\item{several solid objects can be followed simultaneously in the measurement volume.}
\end{enumerate}

This list demonstrates the capabilities of this novel measurement technique. Here we applied it to the dynamics of big inertial particles in turbulent flows (see section 6) and identified the  zone of interaction between the particles and the flow.  Other questions that can be addressed are, for example, the study of collisions of particles with equal and  different particle sizes, and the dynamics of non-spherical objects in turbulent  flows. At high particle number density  this technique can also be used to investigate the dynamics of wet granular matter and of the flow between the grains. 

\ack We are very grateful to Haitao Xu for his many important contributions to the techniques used in this investigation.  We thank J. -F. Pinton, A. Pumir and R. Zimmermann for helpful discussions. Support from COST Action MP0806 is kindly acknowledged. This work was funded generously by the Max Planck Society   and the Marie Curie Fellowship, Program PEOPLE - Call FP7-PEOPLE-IEF-2008  Proposal N$^{o}$ 237521. 

\section*{References}
\bibliographystyle{unsrt}
\bibliography{MST_BigPart}

\end{document}